\documentstyle[12pt]{article}
\input{epsfig.sty}
\newcommand{\beq}{\begin{eqnarray}}
\newcommand{\eeq}{\end{eqnarray}}
\begin{document}
\title{Review of Recent Neutrino Physics Research}
\author{{\bf Leonard S. Kisslinger}\\
{\bf Physics Department, Carnegie Mellon University, Pittsburgh PA15213}}
\date{}
\maketitle
\noindent
PACS Indices:12.38.Aw,13.60.Le,14.40.Lb,14.40Nd
\vspace{1mm}
\begin{abstract}
  We review recent research in neutrino physics, including neutrino oscillations
to test time reversal and CP symmetry violations, the measurement of parameters
in the U matrix, sterile neutrino emission causing pulsar kicks, and neutrino
energies in the neutrinosphere.
\newline
\end{abstract}

\Large

\section{Table of Contents}
\vspace{2mm}

{\bf The Topics Reviewed Are:}
\vspace{2mm}

{\bf Quantum States, Symmetries, Neutrino Oscillations}
\vspace{2mm}

{\bf Time Reversal Violation (TRV) for Various Energies}

\hspace{1cm}{\bf and Baselines}
\vspace{2mm}

{\bf A Proposed TRV experiment}
\vspace{2mm}

{\bf CP Violation (CPV) for Various Energies }

\hspace{1cm}{\bf and Baselines}
\vspace{2mm}

{\bf The LBNE Projct, CP and $\delta_{CP}$}
\vspace{2mm}

{\bf  Extracting $\theta_{13}$ Via Reactor Experiments}
\vspace{2mm}

{\bf New Experimental and Theoretical Results }

\hspace{1cm}{\bf for Sterile Neutrinos and Pulsar Kicks}
\vspace{2mm}

{\bf Neutrino Effective Masses in a Neutrinosphere}
\newpage
\normalsize

\section{HAMILTONIAN, QUANTUM STATES, ENERGY EIGENSTATES, 
SYMMETRIES}

 We now review some basic quantum theory needed to obtain the time 
dependence of energy eigenstates, needed for neutrino oscillations and 
discrete symmetries, which are tested by neutrino oscillations.

\subsection{Hamiltonian and Energy Eigenstates}

In quantum theory one deals with states and operators. A quantum state 
represents a system, and a quantum operator operates on a state.

  The Hamiltonian = $H$ is an operator. Consider a state $A$ in space-time:
 $|A(\vec{r},t)>$,
\beq
\label{H}
      H|A(\vec{r},t)>&=&i(\partial/\partial t)|A(\vec{r},t)> \; .
\eeq

  If the state $|A>$ is an eigenstate of the operator $A$, $A|A>=a|A>$,
where a is a constant, the value of $A$ in the state $|A>$. An energy
eigenstate is an eigenstate of $H$:
\beq
\label{Eeigen}
    H |A> &=& E_A |A> \; ,
\eeq
where $E_A$=energy, and $|A>$ is an energy eigenstate. In space-time
\beq
\label{Eeigspacetime}
        |A(\vec{r},t) > &=& e^{-iE_A t} |A(\vec{r})> \; ,
\eeq
since  $He^{-iE_A t} |A(\vec{r})> = i(\partial/\partial t)e^{-iE_A t} |A(\vec{r})>
 = E_A e^{-iE_A t} |A(\vec{r})>$.
\subsection{Discrete Symmetries:}

The discrete symmetries that we deal with are parity, charge conjugation, and
time reversal, with operators $P,C,{\rm \;and\;}T$.

  The parity operator is defined by
\beq
\label{P}
    P|A(\vec{r},t)>&=& \eta_p |A(-\vec{r},t)> \; ,
\eeq
with $|\eta_p|=1$. The charge conjugation operator is defined by
\beq
\label{C}
  C|{\rm particle}>&=&\eta_c|\rm anti-particle>\; ,
\eeq
with $|\eta_c|=1$. 

If the Hamiltonian is invariant under time reversal, $ T H T^\dagger = H$,
then
\beq
\label{T}
         T e^{-iH(t_2-t_1)}T^\dagger &=&e^{-iH(t_1-t_2)} \; .
\eeq

The CPT Theorem: If the Lagrangian is local, $L=L(x^\mu)$, and invariant
to Lorentz Transformations CPT is conserved, so TRV =CPV in magnitude,
where TRV is time reversal violation and CPV is CP (operator C $\times$
operator P) violation.

 T and CP violations in neutrino oscillations have long been of interest,
see Ref\cite{ahlo01} for definitions and notation. For matter effects see
Ref\cite{jo04}.

\subsection{Neutrino Oscillations}

 Neutrinos are produced (e.g. by proton-proton collisions) with a flavor,
the three flavors being electron, muon, tau neutrinos. They have no definite
mass. As we now show, this leads to Neutrino Oscillations. We use units for
which c=h=1, with c= the speed of light and h=Plank's constant.

A neutrino with mass $m_i$ at rest ($E_i=m_i$) has t-dependence $ |\nu_i,t> = 
e^{-im_i t}|\nu_i>$.  A neutrino with flavor a is related to the mass neutrino 
eigenstates, $\alpha$=1,2,3 by
\beq
\nu_a &=& U \nu_\alpha \nonumber \; ,
\eeq
{\bf with the unitary transformation, U}
\large
\beq
 U=\left( \begin{array}{lcr} c_{12}c_{13} &s_{12}c_{13} & s_{13}
e^{-i \delta_{CP}} \\
     -s_{12}c_{23}-c_{12}s_{23}s_{13}e^{i\delta_{CP}} & 4c_{12}c_{23}-s_{12}
s_{23}s_{13}e^{i\delta_{CP}} & s_{23}c_{13} \\
s_{12}s_{23}-c_{12}c_{23}s_{13}e^{i\delta_{CP}} & -c_{12}s_{23}-s_{12}c_{23}
s_{13}e^{i\delta_{CP}} & c_{23}c_{13} \end{array} \right) \nonumber \; .
\eeq
\normalsize
The parameters are angles $\theta_{ij}$, (i,j) = (1,2), (1,3), (23),
with $s_{ij}\equiv sin(\theta_{ij}),c_{ij}\equiv cos(\theta_{ij})$; and the
angle $\delta_{CP}$.

Therefore, the electron neutrino state is related to the mass neutrino states
by $|\nu_e> = \sum_i U_{ei} |\nu_i>$ and  the time dependence of electron
 neutrinos at rest isc $|\nu_e,t> = \sum_i U_{ei} |\nu_i,t> = \sum_i U_{ei} 
e^{-im_i t}|\nu_i>$  Since the three neutrino masses are different, one finds
 $|\nu_e,t> = c_e(t) |\nu_e>+ c_\mu(t) |\nu_\mu>+c_\tau(t) |\nu_\tau>$,
 where $c_f(t)$ give the amplitude of flavor f for time t. There are similar
relations for the mu and tau neutrinos.
\vspace{3mm}

 From this one sees that as neutrinos travel they oscillate into neutrinos 
with a different flavor.

\section{T, CP, CPT Violation and Neutrino Oscillations}

 Defining $P(\nu_\alpha\rightarrow \nu_\beta)$ = transition probability
of flavor $\alpha$ to flavor $\beta$ neutrino, the T, CP, and CPT violating
probability differences are defined as (with $\bar{\nu}$ an anti-$\nu$:
\beq
        \Delta P^T_{\alpha \beta} &=&P(\nu_\alpha\rightarrow \nu_\beta)-
P(\nu_\beta\rightarrow \nu_\alpha) \nonumber \\
        \Delta P^{CP}_{\alpha \beta} &=&P(\nu_\alpha\rightarrow \nu_\beta)-
P(\bar{\nu}_\alpha\rightarrow \bar{\nu}_\beta) \nonumber \\
        \Delta P^{CPT}_{\alpha \beta} &=&P(\nu_\alpha\rightarrow \nu_\beta)-
P(\bar{\nu}_\beta\rightarrow \bar{\nu}_\alpha) \nonumber \; ,
\eeq
where $\alpha,\beta = e, \mu, \tau$

 The time evolution matrix, $S(t,t_0)$, is used to derive $\Delta P^T_{ab}$:
\beq
             |\nu(t)> &=& S(t,t_0)|\nu(t_0)> \nonumber \\
             i\frac{d}{dt}S(t,t_0) &=& H(t) S(t,t_0) \nonumber \; ,
\eeq
In the vacuum
\beq
        S_{ab}(t,t_0)&=& \sum_{j=1}^{3} U_{aj} exp^{i E_j (t-t_0)} U^*_{bj}
\eeq

 The neutrino-electron potential for neutrinos traveling
 through matter for matter density $n_e$ =3 gm/cc is
\beq
 V &=& \sqrt{2} G_F n_e=1.13 \times 10^{-13} {\rm \; eV} \nonumber 
\eeq
where $G_F$ is the universal weak interaction Fermi constant.
Note $P(\nu_\alpha\rightarrow \nu_\beta) =  |S_{\beta \alpha}|^2$; and
 $P(\bar{\nu}_\alpha\rightarrow \bar{\nu}_\beta) =  |\bar{S}_{\beta \alpha}|^2$
with $V \rightarrow -V$. I.e., $V$(anti-neutrinos)=-$V$(neutrinos).

\subsection{Time Reversal in Neutrino Oscillations}

 This discussion of TRV is based on Ref\cite{hjk11}, in which the formalism
of Freund\cite{freund01} is used. 

 The TRV electron-muon probability difference in matter is
\beq
  \Delta P^T_{e \mu}&=& |S_{21}|^2-|S_{12}|^2 \nonumber \\
   S_{12} &=& c_{23} \beta -is_{23} a A_a \nonumber \\
     S_{21} &=& -(c_{23} \beta +is_{23} a C_a) \nonumber \\
      a &=& s_{13}(\Delta -s_{12} \delta) \nonumber \\
    \Delta P^T_{e \mu}&=&- 2 s_{13}s_{23}c_{23}(\Delta-s_{12}\delta)
Im[e^{-i \delta_{CP}}\beta^*(A_a-C_a^*)] \; ,
\eeq
with
\beq 
     A_a & \simeq & f(t,t_0) I_\alpha*(t,t_0) \\
       I_\alpha*(t,t_0)&=& \int_{t_0}^t dt' \alpha^*(t',t)f(t',t) \nonumber \\
     \alpha(t,t_0) &=& cos\omega (t-t_0) -i sin (2\theta) sin \omega (t-t_0)
 \nonumber \\
          f(t,to) &=& e^{-i \Delta (t-t_0)}   \nonumber \\
 2 \omega &=& \sqrt{\delta^2 + V^2 -2 \delta V cos(2 \theta_{12})} \nonumber \\
         \beta &=& -i sin2\theta sin\omega L \nonumber \\
        C_a &=& A_a \nonumber \\
    cos(2\theta) &=& \frac{\delta cos(2 \theta_{12}) -V}{2 \omega} 
 \nonumber \;.
\eeq

We choose $s_{13}=.187$, $s_{23}=c_{23}=.707$,
$\theta_{12}=32^o$. Note $\delta =\delta m_{12}^2/(2 E) \ll\Delta = 
\delta m_{13}^2/(2 E)$ From $ \Delta P^T_{e \mu}= |S_{21}|^2-|S_{12}|^2$ it follows
 that
\beq
\label{deltaT}
     \Delta P^T_{e \mu}&=&- 2 s_{13}s_{23}c_{23}(\Delta-s_{12}\delta)
Im[e^{-i \delta_{CP}}\beta^*(A_a-C_a^*)] \nonumber \\
   A_a-C_a^*&= & 2iIm[A_a] \simeq i\frac{2}{\Delta}[cos\Delta L- cos\omega L] 
 {\rm  \;,\;therefore,} \\
  \Delta P^T_{e \mu}&\simeq&-4 s_{13}s_{23}c_{23}sin\omega L
sin2\theta(cos\Delta L-cos\omega L) \nonumber \\
    &\simeq& 0.374 sin\omega L sin2\theta (cos\omega L-cos\Delta L) 
\nonumber \; .
\eeq 
\clearpage

 $\Delta P^T_{e \mu}$ is shown as a function of L for E=1 , a function 
of E for L=735 km
\vspace{3cm}
\begin{figure}[ht]
\begin{center} 
\epsfig{file=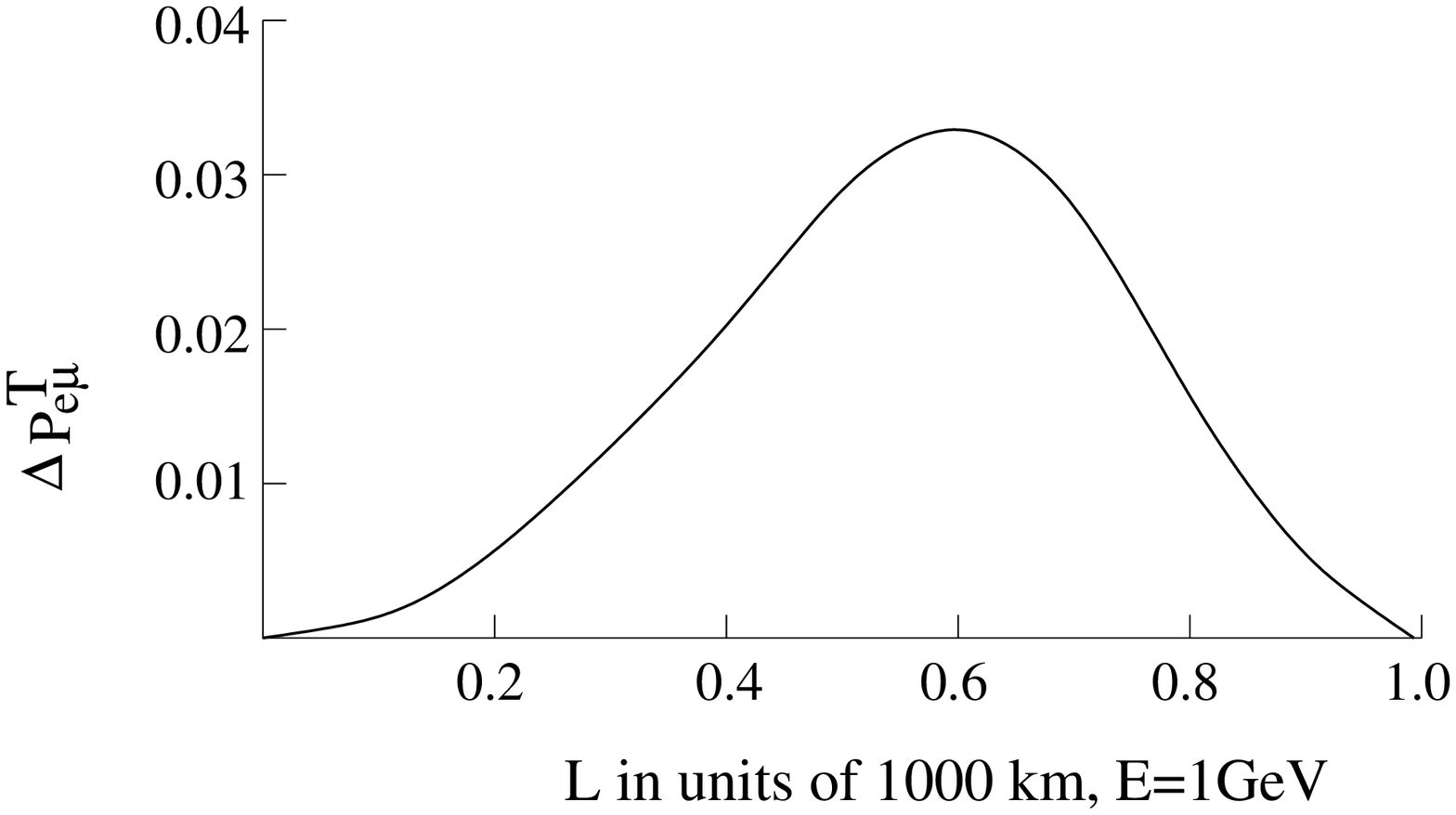,height=4.0 cm,width=12 cm}
\end{center}
\end{figure}
\vspace{2cm}

\begin{figure}[ht]
\begin{center}
\epsfig{file=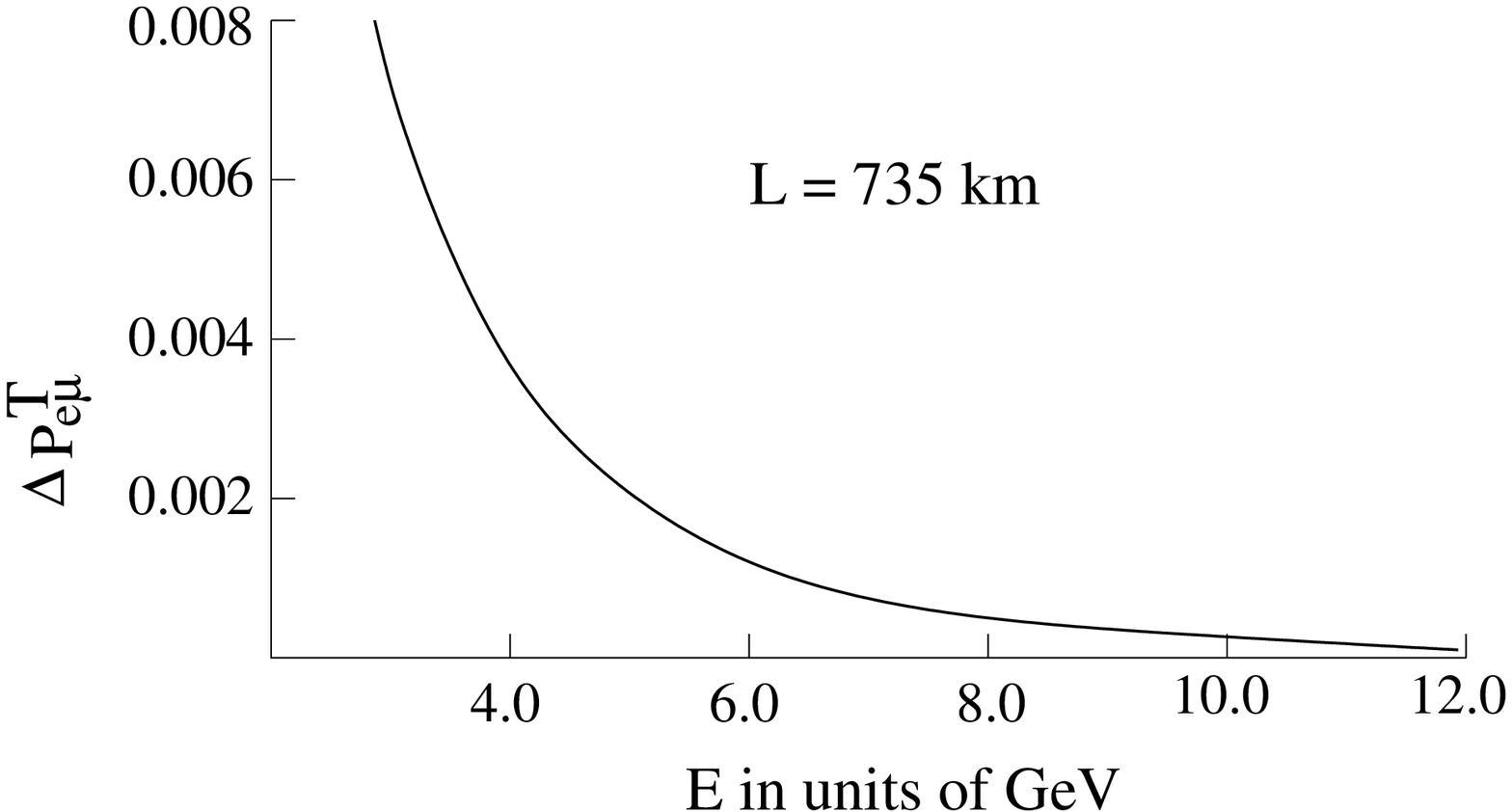,height=4.0cm,width=12 cm}
\end{center}
\end{figure}

 Note that for E= 1 GeV and L=735km (Minos parameters)
TRV is approxmately 3\%, which could be measured if muon as well as electron
neutrino beams were available. Unfortunately, this is not now possible.

\newpage
\subsection{Proposed TRV Experiment}

Since none of the neutrino oscillation facilities have beams of both 
$\nu_e$ and $\nu_\mu$ beams, with $\Delta P^T_{e\mu}=
 P(\nu_e \rightarrow \nu_\mu)-P(\nu_\mu \rightarrow \nu_e)$ the time reversal 
violation (TRV) probability, it is not possible to measure TRV at the present 
time.
\vspace{1cm}
  Recently a TRV experiment was proposed\cite{khj12}.
The proposed experiment is shown in Fig. 1.  The neutrino beam
is aimed at a new detector at the surface of the earth 735 km past the
Soudan mine (position 3 in the figure), only a small deviation from the
present MINOS neutrino beam (position 1). There would be a 1\% increase 
in the electron neutrino probability that 
one would obtain with the 10\% muon neutrino flux at the position of the 
Soudan mine. This would be a means for the measurement of TRV.
Soudan mine. This would be a means for the measurement of TRV.

\begin{figure}[ht]
\begin{center}
\epsfig{file=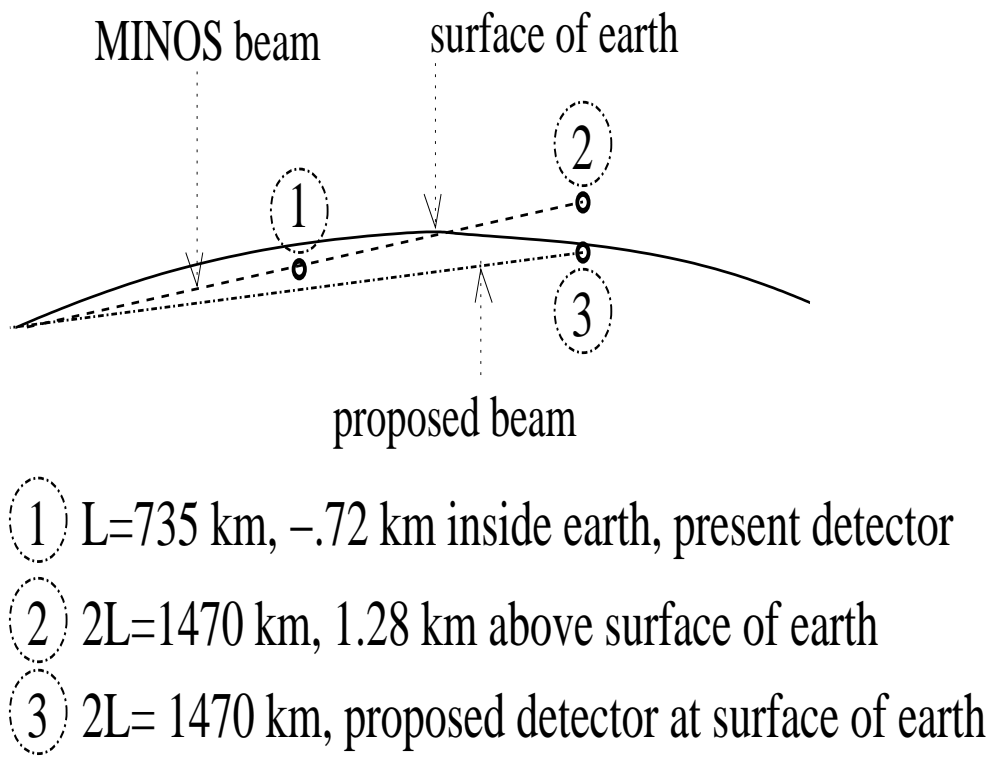,height=10.0cm,width=14cm}
\end{center}
\caption{The present Minos neutrino beam and a proposed beam}
\end{figure}
\clearpage
\subsection{CPV via Neutrino Oscillations}

We start with an overview of CP and CV Violation (CPV)
\vspace{5mm}

CPV has a long history:
\vspace{3mm}

\hspace{5mm} $K^0_L \rightarrow \pi^+ + \pi^{-}$\cite{ccft64}
\vspace{3mm}

\hspace{5mm} $K^0_L \rightarrow 2 \pi^0$\cite{ckrw67}

\hspace{5mm} $K^0_L \rightarrow \pi^0 + \nu + \bar{\nu}$\cite{ll89}
\vspace{5mm}

The first study in the present work is an estimate the 
$\nu_\mu$ to $\nu_e$ conversion probability using parameters for the
baseline and energy corresponding to MiniBooNE, JHF-Kamioka, 
MINOS, and CHOOZ-Double Cho0z, which are 
on-going projects, although the CHOOZ project does not have a beam of 
$\nu_\mu$ neutrinos.The two main parameters of interest in the present work 
are $\delta_{CP}$, which is essentially unknown, and $\theta_{13}$.

 The LBNE Project, where neutrino beams produced at Fermilab
with a baseline of L $\simeq$ 1200 km would be detected with deep underground 
detectors, has been proposed for studying CPV and the $\delta_{CP}$ parameter.
We discuss the recent work on this project.

 Next are shown results for $\mathcal{P}(\nu_\mu \rightarrow \nu_e)$ 
for MiniBooNE, JHF-Kamioka, MINOS, Chooz parameters as a guide for future
CPV experiments.

 The angle $\theta_{13}$ will be measured by the Daya Bay experiment in 
China, the Double Chooz project in France, and RENO in Korea,
via $\bar{\nu}_e$ disappearance.  $\bar{\nu}_e$ disappearance is derived
using the recent result from the Daya Bay project that 
$s_{13} \simeq 0.15$ as a test of the equation used by the Daya Bay experimental
group compared to the more modern S-Matrix theory. It is shown that for some
baselines it would be important to use the more modern theory.

 Finally, using the expected range of values for $\theta_{13}$, 
CPV is estimated for $\mu-e$ neutrino oscillation for the entire range of 
$\delta_{CP}$ to help in the planning for future CPV experiments.
\vspace{5mm}
\subsection{CP VIOLATION AND THE LBNE PROJECT}

 The two main objectives of the LBNE Project are to measure the 
$\delta_{CP}$ parameter and CPV via neutrino oscillations. 
\vspace{5mm}

 A study of CPV using the baseline (L=1200 km) and energies expected 
for the future LBNE Project, calculating CPV as a function of 
$\delta_{CP}$ was carried out\cite{lsk12} to help in the design of the 
project. An essential aspect of the determination 
of CPV is the interaction of neutrinos with matter as they travel along the 
baseline.

 The CPV probability differences (note that the C operator changes a
particle to its antiparticle) are defined as $\Delta P^{CP}_{ab}$
\beq
   \Delta P^{CP}_{ab}&=& P(\nu_a \rightarrow \nu_b)
-P(\bar{\nu}_a \rightarrow \bar{\nu}_b) \nonumber \; .
\eeq
\newpage
 In our present work we study $ P(\nu_\mu \rightarrow \nu_e)$ and
 $P(\bar{\nu}_\mu \rightarrow \bar{\nu}_e)$, since the neutrino 
beams at MiniBooNE,  JHF-Kamaoka, MINOS, and LBNE, as well as most
 other experimental facilities, are muon or anti-muon neutrinos.

 The probability of CPV for $\mu, e$ neutrinos, 
$\Delta\mathcal{P}^{CP}_{\mu e}$ is
\beq
\label{CPV}
  \Delta P^{CP}_{\mu e} &=& P(\nu_\mu \rightarrow \nu_e)
-P(\bar{\nu}_\mu \rightarrow \bar{\nu}_e) \nonumber \\
         &=&  |S_{12}|^2- |\bar{S}_{12}|^2 \nonumber \\
        S_{12} &=& c_{23} \beta -is_{23} a e^{-i\delta_{CP}} A_a \nonumber \\
  \bar{S}_{12} &=&  c_{23} \bar{\beta} -is_{23} a e^{i\delta_{CP}} \bar{A_a}
\nonumber \; ,
\eeq
with $ c_{23}, s_{23}, \beta, A_a$ defined above.
We use $s_{12} =0.56$ and $s_{23}=0.707$; and choose $s_{13}=0.19$, which
one of the past projects found. Using the matter density $\rho$=3 gm/cc,
$V =1.13 \times 10^{-13}$. Note that for antineutrinos $\delta_{CP} 
\rightarrow -\delta_{CP}$. $\bar{\beta}= \beta (V \rightarrow -V)$ and 
$\bar{A}= A(V \rightarrow -V)$.

Note that if $V=0$, (vacuum), CPT is always conserved for normal theories,
but for $V=\neq 0$ CP an T symmetries are independent

 Using conservation of probabiltiy, $|A_a|^2=|\bar{A_a}|^2$, 
one finds (we do not give the rather complicated result here. See 
Ref\cite{lsk12}).

\beq
\label{DCPV}
  \Delta P^{CP}_{\mu e} &=& c_{23}^2(|\beta|^2-|\bar{\beta}|^2)
-2 c_{23} s_{23} a Im[\beta e^{-i\delta_{CP}}A^*- e^{i\delta_{CP}} 
\bar{\beta} \bar{A}^*] \nonumber \; .
\eeq 

  The estimates of $\Delta P^{CP}_{\mu e}$ for the baseline and energies
expected for the LBNE Project are shown in Fig. 2.
\subsection{CONCLUSIONS FOR LBNE PROPOSED PROJECT:}
\vspace{5mm}

 In the LBNE Report (V. Barger $et\;al$, Report of the US long baseline 
neutrino experiment study (arXiv:0705.4396) results of extensive studies have 
shown that future experiments can extend our knowledge of neutrino oscillations
beyond present and planned experiments. Since there will be both $\nu_\mu$ and 
$\bar{\nu}_\mu$ beams, the LBNE Project can test CPV.

 We have estimated CP violation for the LBNE Project, with a baseline 
L=1200 km as a function of $\delta_{CP}$ for $\delta_{CP}$ = 0 to $\pi/2$ 
for energies of 1, 2, and 3 GeV. CPV over 3\% was found
with $\delta_{CP}=\pi/2$ for some  energies, which the LBNE Project should 
be able to measure. Even for $\delta_{CP}=0$, for which CPV is entirely a 
matter effect, CPV probabilities of over 1\%  were found for E= 1 GeV, so 
the LBNE Project should be able to measure CPV for any expected values of 
$\delta_{CP}$.

 We believe that these calculations should be useful in planning the future
LBNE Project.
\clearpage

  $\Delta P^{CP}_{\mu e}(E,\delta_{CP})$ as a
function of $\delta_{CP}$ for energies E= 1, 2, 3 GeV, are shown in 
the figure.
\vspace{6cm}

\begin{figure}[ht]
\begin{center}
\epsfig{file=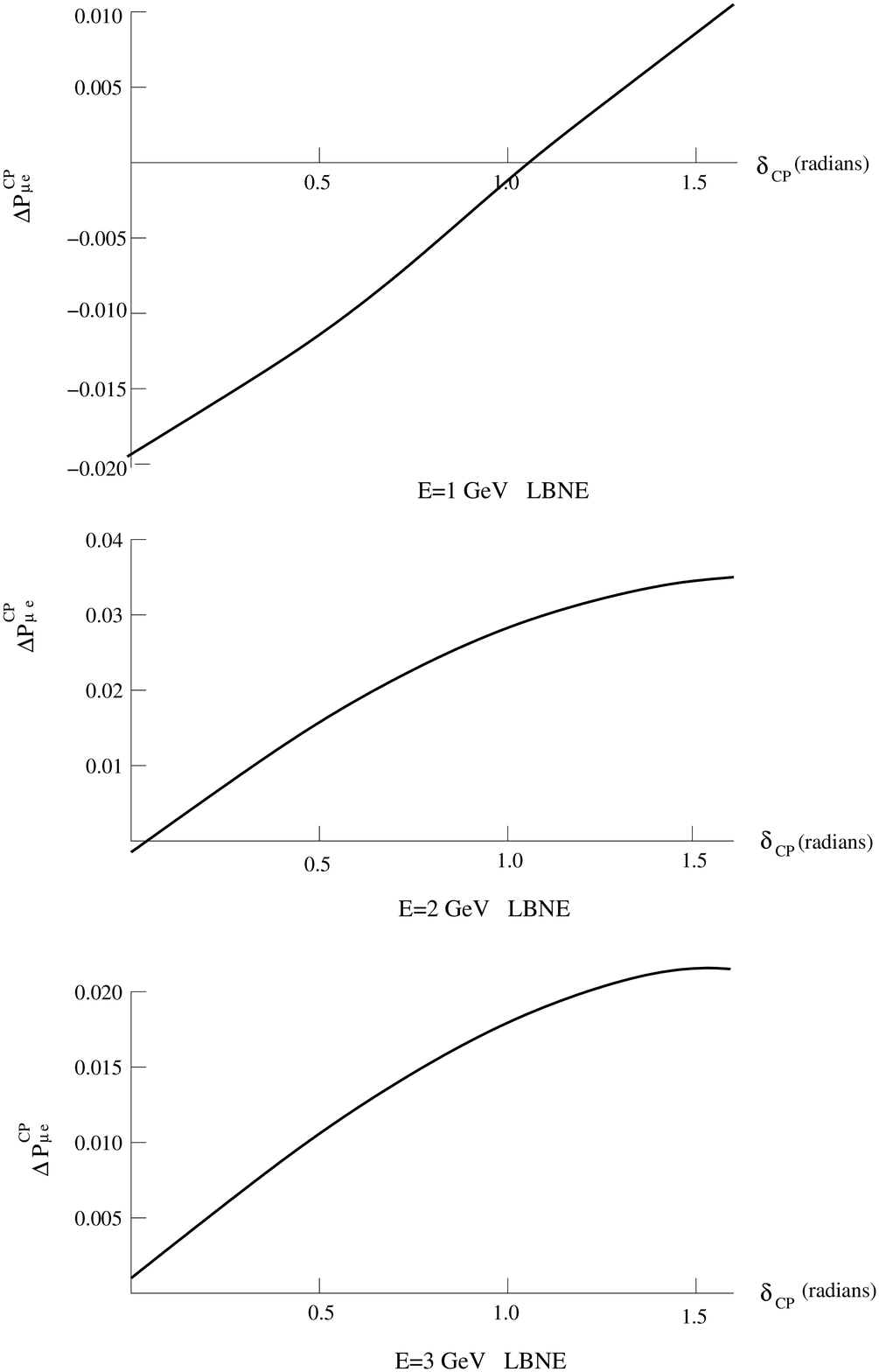,height=12cm,width=12cm}
\end{center}
\caption{The ordinate is $\Delta \mathcal{P}(\nu_\mu \rightarrow\nu_e)$ for 
LBNE(L=1200 km), Energy=E in GeV, as a function of $\delta_{CP}$}
\end{figure}

\newpage

\subsection{CPV via Neutrino Oscillation in Matter and Parameters 
$s_{13},\delta_{CP}$}

 [L.S. Kisslinger, E.M. Henley, M.B. Johnson; arXiv:1203.6613/hep-ph;
 Int. J. of Mod. Phys. E 21, 1250065 (2012)]
\vspace{3mm}

 An estimate of the dependence of $\nu_{\mu}$ to $\nu_{e}$ 
conversion on parameters $\theta_{13}$ and $\delta_{CP}$ for experimental 
facilities studying neutrino oscillations was carried out\cite{khjz12}.
Of particular interest is the dependence on $\delta_{CP}$

 Reactor experiments at Daya Bay, Double Chooz, and RENO
measuring $\bar{\nu_e}$ disappearance are in progress, with the objective 
of determining $\theta_{13}$, an essential parameter for neutrino oscillations,
to 1\%

 The S-Matrix theory is used to estimate $\bar{\nu_e}$ 
disappearance and compare estimates based on an older theory being used 
by Daya Bay, etc., experimentalists to  extract $\theta_{13}$, using values 
of $\theta_{13}$ within known limits to
estimate the dependence of $\nu_{\mu}$ - $\nu_{e}$ CPV probability on 
$\delta_{CP}$ in order to suggest new experiments to measure CPV for 
neutrinos moving in matter.
\vspace{3mm}

 The transition probability $P(\nu_\mu \rightarrow\nu_e)$
is obtained from $S_{12}$:
\beq
\label{PueS12}
 P(\nu_\mu \rightarrow\nu_e) &=& |S_{12}|^2=Re[S_{12}]^2+
Im[S_{12}]^2 \nonumber \; ,
\eeq
with $S_{12} = c_{23} \beta -is_{23} a e^{-i\delta_{CP}} A_a$.
 Since $\delta = \delta m_{12}^2/(2 E)$ and $\Delta = \delta m_{13}^2/(2 E)$,
$\delta \ll \Delta$; and one can show 
that
\beq
\label{Pue}
         Re[S_{12}] &=& s_{23}a[cos(\bar{\Delta}L+\delta_{CP})Im[I_{\alpha*}]
-sin(\bar{\Delta}L+\delta_{CP}) Re[I_{\alpha*}] \nonumber \\
         Im[S_{12}] &=& -c_{23}sin2\theta sin\omega L -s_{23} a
[cos(\bar{\Delta}L+\delta_{CP}) Re[I_{\alpha*}] \nonumber \\
        &&+sin(\bar{\Delta}L+\delta_{CP}) Im[I_{\alpha*}]] \nonumber \; ,
\eeq
and $ Re[I_{\alpha*}] \simeq
 \frac{sin \bar{\Delta} L}{\bar{\Delta}};  Im[I_{\alpha*}] \simeq 
\frac{1-cos \bar{\Delta} L}{\bar{\Delta}}$.

 From this one obtains the mu to e neutrino oscillation probability
\beq
\label{CPue}
    P(\nu_\mu \rightarrow\nu_e) &\simeq& (c_{23}s_{12}c_{12}
(\delta/\omega) sin\omega L)^2 +2(s_{23} s_{13})^2(1- cos\bar{\Delta} L)
\nonumber \\
  && +2s_{13}s_{12}c_{12}s_{23}c_{23} (\delta/\omega) sin\omega L
\nonumber \\
  && (cos(\bar{\Delta}L+\delta_{CP})sin\bar{\Delta}L+
sin(\bar{\Delta}L+\delta_{CP})(1-cos\bar{\Delta}L)) \nonumber \;.
\eeq

 Both $s_{13}$=.19 and .095 were used to show the effect of $s_{13}$.
\vspace{3mm}

  The results for $P(\nu_\mu \rightarrow\nu_e)$ are shown in 
Fig.3. These results can provide guidance for future 
experiments on CPV via $\nu_\mu \leftrightarrow \nu_e$ oscillation. 

 We calculated $P(\nu_\mu \rightarrow\nu_e)$ for $\delta_{CP}$
from -$\pi$/2 to $\pi$/2, and the results are almost independent of 
$\delta_{CP}$. The results for CHOOZ are shown in preparing for the following
subsection on $\bar{\nu_e}$ disappearance, even though Double Chooz, Daya
Bay, and RENO projects have beams of $\bar{\nu}_e$ rather than $\nu_\mu$ 
neutrinos.
\clearpage

\begin{figure}[ht]
\begin{center}
\epsfig{file=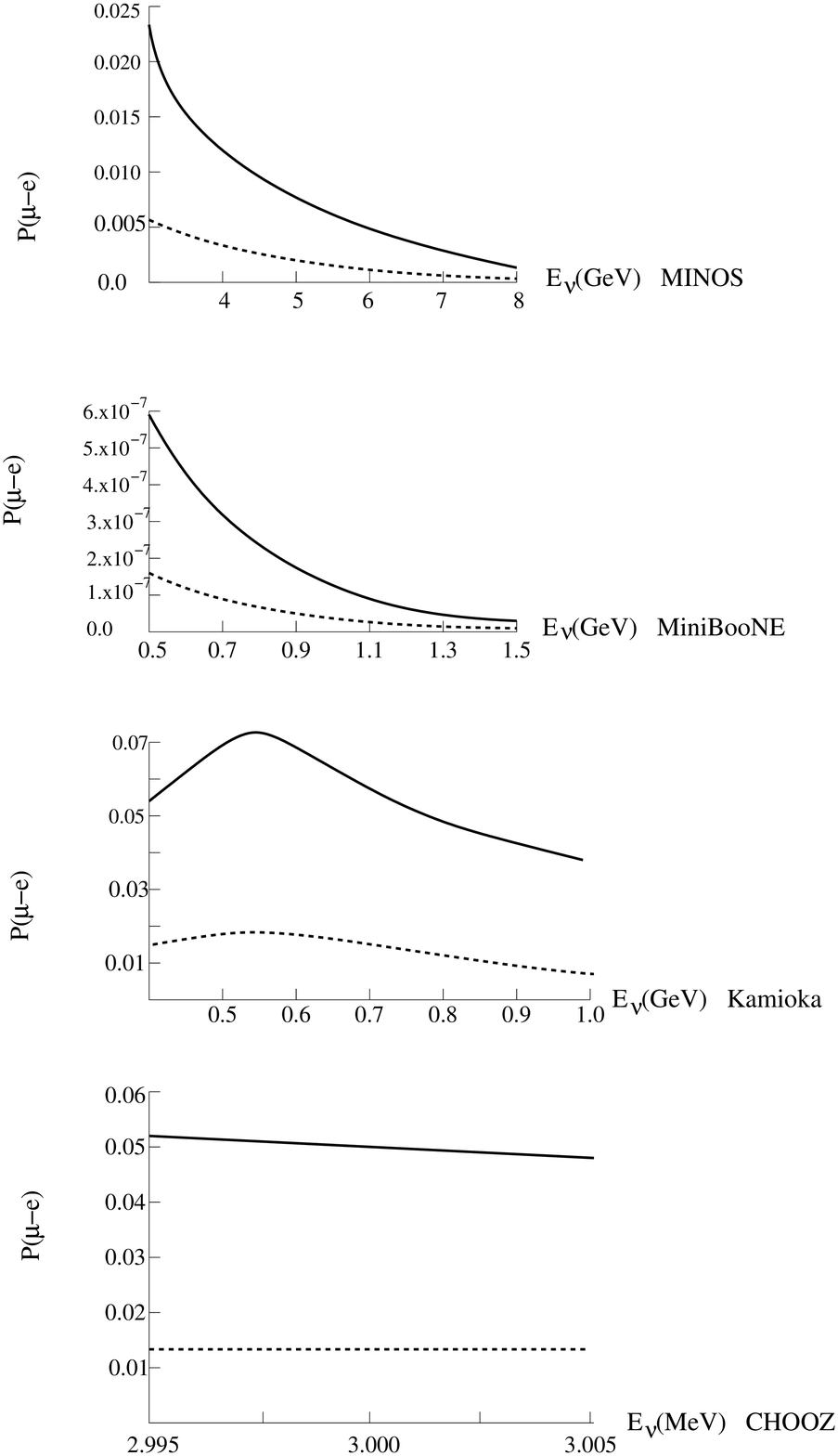,height=18cm,width=12cm}
\end{center}
\caption{\hspace{5mm} The ordinate is $P(\nu_\mu \rightarrow\nu_e)$ 
for MINOS(L=735 km),
 MiniBooNE(L=500m), JHF-Kamioka(L=295 km), and 
CHOOZ(L=1.03 km).
\hspace {5mm} Solid curve for $s_{13}$=.19 and dashed curve for 
$s_{13}$=.095. The curves are almost independent of $\delta_{CP}$.}
\end{figure}
\clearpage

\subsection{ $\bar{\nu_e}$ DISAPPEARANCE DERIVED USING S-MATRIX
THEORY COMPARED  TO Daya Bay EVALUATION:}

\vspace{1cm}

 We derive $\bar{\nu_e}$ disappearance, $P(\bar{\nu}_e \rightarrow 
\bar{\nu}_e)$, defined as
\beq
\label{disappear}
  P(\bar{\nu}_e \rightarrow \bar{\nu}_e)&=& 1-
 P(\bar{\nu}_e \rightarrow \bar{\nu}_\mu)-
 P(\bar{\nu}_e \rightarrow \bar{\nu}_\tau) \nonumber \; ,
\eeq
 using the S-matrix method  and compare it
to the expression for $P(\bar{\nu}_e \rightarrow \bar{\nu}_e)$
used by the Daya Bay, Double Chooz, and RENO.
\vspace{5mm}

 The expression derived decades ago used by Daya Bay and the other
reactor experimentalists to find $\theta_{13}$ from $\bar{\nu_e}$ 
disappearance is
\beq
\label{dc06P}
  P^{DB}(\bar{\nu}_e \rightarrow \bar{\nu}_e)&\simeq & 1-
 4(s_{13}c_{13})^2 sin^2(\frac{\Delta L}{2}) \nonumber \; ,
\eeq
where $\Delta \equiv  \delta m_{13}^2/(2 E)$, and $s_{13},c_{13}=
sin\theta_{13},cos\theta_{13}$.
\vspace{3mm}

 In the S-matrix method the probability of $\bar{\nu}_e$ oscillation 
to $\bar{\nu}_\mu$ and $\bar{\nu}_\tau$ is given by
\beq
\label{barCP} 
 P^{SM}(\bar{\nu}_e \rightarrow \bar{\nu}_\mu)&=&|\bar{S}_{21}|^2 
\nonumber \\
 P^{SM}(\bar{\nu}_e \rightarrow \bar{\nu}_\tau)&=&|\bar{S}_{31}|^2 
\nonumber \; .
\eeq

 We take $\delta_{CP}=0$, since as was mentoned the relevant oscillation
probablities are essentially independent of $\delta_{CP}$.
Therefore $|\bar{S}_{21}|^2$=$|S_{12}(V\rightarrow -V)|^2$,
and $|\bar{S}_{31}|^2$=$|S_{12}|^2(V\rightarrow -V,c_{23}\rightarrow s_{23},
 s_{23}\rightarrow -c_{23}$). 
\vspace{3mm}

  Note that $s_{23}^2\simeq c_{23}^2\simeq 1/2$.
\vspace{3mm}

 From the above, the result for the S-matrix theory of anti-electron 
disappearance is

\beq
\label{barSM}
 P^{SM}(\bar{\nu}_e \rightarrow \bar{\nu}_e)&=& 1-
[(.46 \delta sin\bar{\omega} L/\bar{\omega})^2 +2(s_{13})^2(1- 
cos\bar{\bar{\Delta}} L)] \nonumber
\; ,
\eeq
with $\bar{\bar{\Delta}}=\Delta +(V-\delta)/2$, 2 $\bar{\omega}=
 \sqrt{\delta^2 + V^2 +2 \delta V cos(2 \theta_{12})}$, and 
$\delta=\delta m_{12}^2/(2 E)$
\vspace{5mm}

 Fig. 4 shows $ P(\bar{\nu}_e \rightarrow \bar{\nu}_e)$ for L=1.9 km, the
Daya Bay baseline, for the DB and SM calculations.
\clearpage
{\bf Anti-electron neutrino disappearance for SM vs DB}
\vspace{-4cm}

\begin{figure}[ht]
\begin{center}
\epsfig{file=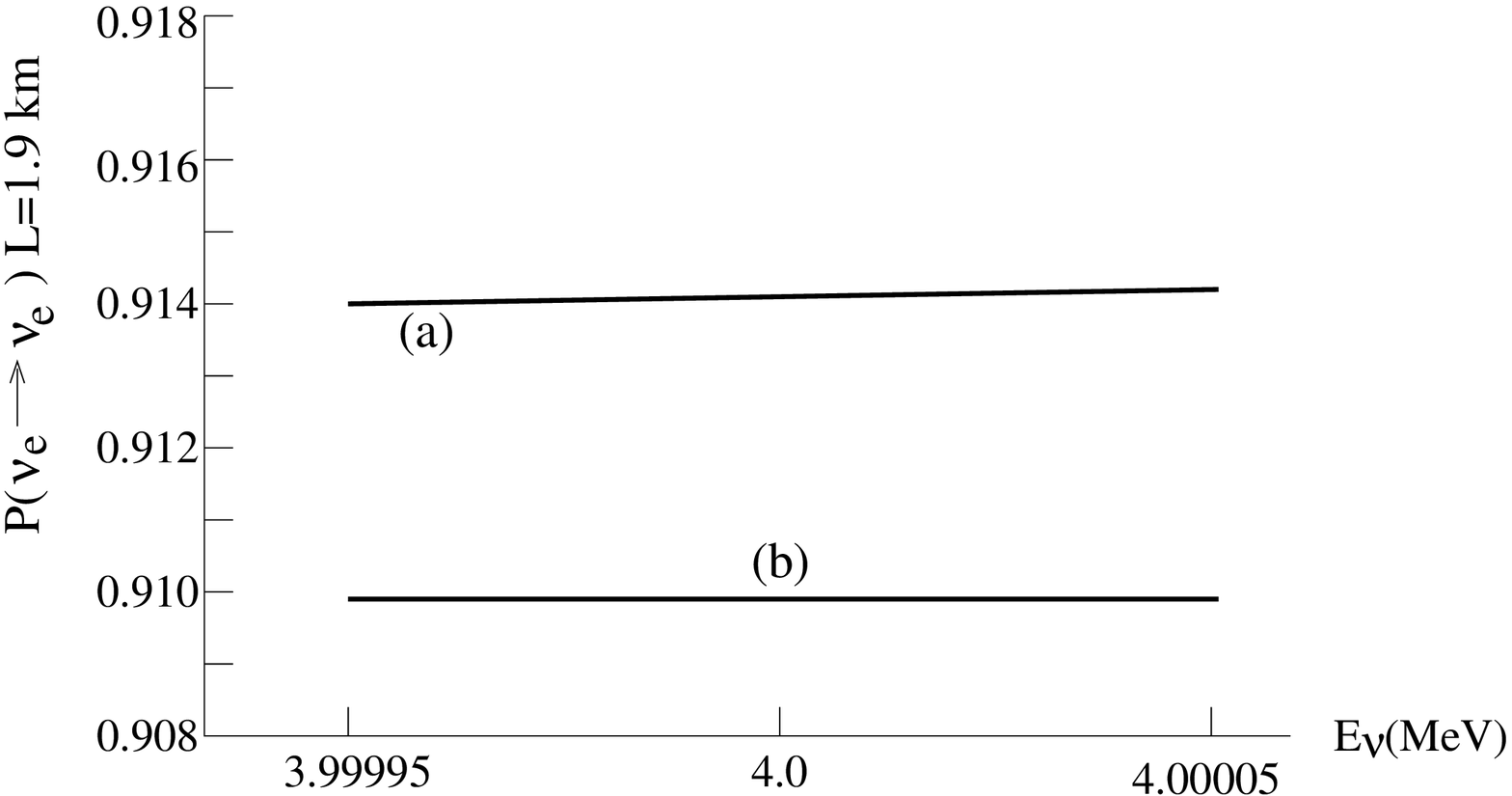,height=12cm,width=14cm}
\end{center}
\caption{$P(\bar{\nu}_e \rightarrow \bar{\nu}_e)$. 
For $s_{13}=.15$ (a) $P^{DB}$ and (b) $P^{SM}$}
\end{figure}
\vspace{1cm}

 From Fig. 4 the ratios $R1, R2$, of $1-P(\bar{\nu}_e \rightarrow 
\bar{\nu}_e)$ for $P^{DB}$ to $P^{SM}$ 
for $s_{13}=.15$, and for $s_{13}=.15$ for $P^{DB}$ and $s_{13}=.147$
for $P^{SM}$,  for E $\simeq$ 4.0MeV and L=1.9 km are

\beq
\label{ratio1}
   R1=\frac{1-P^{DB}(s_{13}=.15)}{1-P^{SM}(s_{13}=.15)}
&=& 1.04 \nonumber \\
   R2=\frac{1-P^{DB}(s_{13}=.15)}{1-P^{SM}(s_{13}=.147)}
&=& 1.00  \nonumber\;,
\eeq
which demonstrates that using the S-Matrix formulation for L=1.9 km and 
E$\simeq$ 4.0 MeV one would extract $s_{13}=.147$ from the data for which
the older formalism finds $s_{13}=.15$. This is a 2\% correction.
\vspace{1cm}

 In Fig. 5 the same calculation is shown for a baseline L=10 km,
as future projects might use a longer baseline for a larger effect given 
$s_{13}$ 
\clearpage

\begin{figure}[ht]
\begin{center}
\epsfig{file=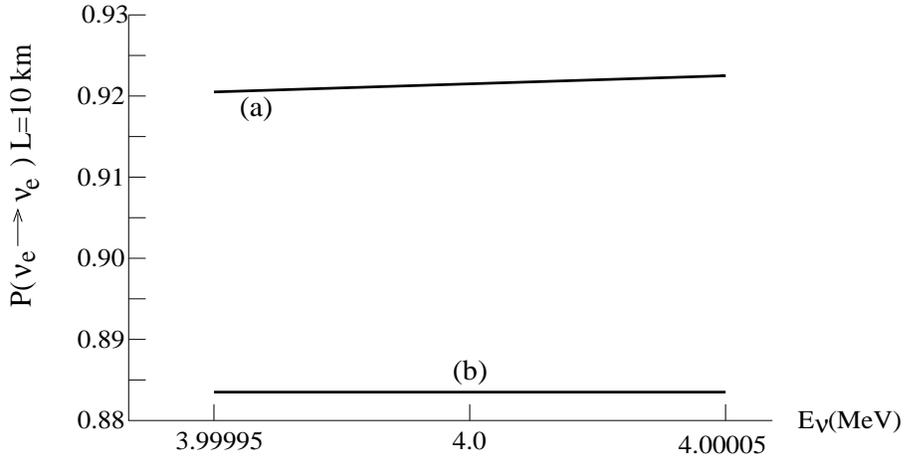,height=6cm,width=12cm}
\end{center}
\caption{$P(\bar{\nu}_e \rightarrow \bar{\nu}_e)$. 
For $s_{13}=.15$ (a) $P^{DB}$ and (b) $P^{SM}$}
\end{figure}

 For E $\simeq$ 4.0MeV and L=10 km the ratios are
\beq
\label{ratio2}
   R1=\frac{1-P^{DB}(s_{13}=.15)}{1-P^{SM}(s_{13}=.15)}
&=& 1.47 \nonumber \\
   R2=\frac{1-P^{DB}(s_{13}=.15)}{1-P^{SM}(s_{13}=.095)}
&=& 1.00 \nonumber \;.
\eeq

 Thus using the S-Matrix formulation for L=10 km and 
E$\simeq$ 4.0 MeV one would extract $s_{13}=.095$ from the data for which
the older formalism finds $s_{13}=.15$. This is a 35\% correction.
\vspace{5mm}

 We have carried out similar calculations for the 
T2K project, with E=0.6 GeV, L=295 km. With both a larger L and 
larger E than Daya Bay,  we find a correction of 2.4\%.

 It is also important to note that our SM method
gives $P(\bar{\nu}_e\rightarrow \bar{\nu}_e)\neq 1.0$ even for
$s_{13}$=0, in contrast to the older DB method
\vspace{5mm}
\Large

{\bf CP Violation: $\Delta P^{CP}_{\mu e}$}
\normalsize

 We now extend the derivation of the transition
probability $P(\nu_\mu \rightarrow \nu_e)$ shown above
to derive the CPV probability:
\beq
\label{CPV2}
  \Delta P^{CP}_{\mu e} &=& P(\nu_\mu \rightarrow \nu_e)
-P(\bar{\nu}_\mu \rightarrow \bar{\nu}_e) \nonumber \\
         &=&  |S_{12}|^2- |\bar{S}_{12}|^2 \nonumber \,
\eeq
with $S_{12}$ defined above and
$  \bar{S}_{12} =  c_{23} \bar{\beta} -is_{23} a e^{i\delta_{CP}} \bar{A_a}$
with $\bar{\beta}= \beta (V \rightarrow -V)$ and $\bar{A_a}= 
A_a(V \rightarrow -V)$. 

\clearpage
 Therefore

\beq
\label{DCPV2}
  \Delta\mathcal{P}^{CP}_{\mu e} &=& c_{23}^2(|\beta|^2-|\bar{\beta}|^2)
-2 c_{23} s_{23} ai (Im[\beta e^{i\delta_{CP}} A^*]
-Im[\bar{\beta}e^{i\delta_{CP}} \bar{A}^*]) \nonumber \; .
\eeq

 From this and our results shown previously
\beq
\label{DCPVf}
&&  \Delta P^{CP}_{\mu e} = c_{23}^2 s_{12}^2 c_{12}^2 \delta^2
(\frac{s^2}{\omega^2}-\frac{\bar{s}^2}{\bar{\omega}^2}) +2 c_{23}s_{23}
s_{12}c_{12}s_{13}\delta (\Delta-\delta s_{12}^2)  \nonumber \\
 &&(sin\theta_{CP}(\frac{s}{\omega}(c-cos\bar{\Delta}L)
\frac{\bar{\Delta}-\omega cos 2\theta}
{\bar{\Delta}^2-\omega^2}+\frac{\bar{s}}{\bar{\omega}}(\bar{c}-
cos\bar{\bar{\Delta}}L)
 \frac{\bar{\bar{\Delta}}-\bar{\omega} cos 2\bar{\theta}}
{\bar{\bar{\Delta}}^2-\bar{\omega}^2})) \nonumber \\
&&- cos\theta_{CP}(\frac{s}{\omega}\frac{sin\bar{\Delta}L(\bar{\Delta}
-\omega cos2\theta)+sin \omega L(\omega+\bar{\Delta}cos 2\theta)}
{\bar{\Delta}^2-\omega^2} \nonumber\\
&&-\frac{\bar{s}}{\bar{\omega}}\frac{sin\bar{\bar{\Delta}}L(\bar{\bar{\Delta}}
-\bar{\omega} cos2\theta)+sin\bar{\omega} L(\bar{\omega}
+\bar{\bar{\Delta}}cos 2\theta)}{\bar{\bar{\Delta}}^2-\bar{\omega}^2})
\nonumber \; .
\eeq

 The results for $\Delta P^{CP}_{\mu e}$ for $s_{13}$=.19 are shown 
in Fig.6. Note that $\Delta\mathcal{P}^{CP}_{\mu e}$ depends strongly on
$\delta_{CP}$, which could lead to a measurement of this parameter. The
large value of $\Delta P^{CP}_{\mu e}$ for CHOOZ is promising
for future experiments. $\Delta P^{CP}_{\mu e}$ is so small (from
about $10^{-10}$ to $10^{-18}$)for MiniBooNE, we do not show the results. 
Similar results for $\Delta P^{CP}_{\mu e}$ for $s_{13}$=.095 are shown 
in Fig.7.
\vspace{5mm}

\Large
{\bf Conclusions for CPV:}
\normalsize

 We have estimated CP violation for a variety of experimental neutrino beam
facilities, for values of the parameter $s_{13}$ =0.19 and .095, and for 
$\delta_{CP}$ from 90 to -90 degrees, since its value is not known. As our 
results show, the probability $P(\nu_\mu \rightarrow\nu_e)$ is 
essentially independent of $\delta_{CP}$,  and therefore the measurement of 
$P(\nu_\mu \rightarrow\nu_e)$ should determine the value of the 
$s_{13}$ parameter.

 Our new results for $\bar{\nu}_e$ disappearance, as is being measured the 
Daya Bay, Double Chooz and RENO projects, make use of a different 
theoretical formulation than that used by these projects to extract $s_{13}$ 
from the data. We have shown that the recent result from the Daya Bay 
collaboration with E$\simeq$4 MeV and L=1.9 km, from which it 
was estimated that $s_{13}\simeq .15$, by our analysis is $s_{13}\simeq .147$, 
a 2\% correction. This is small, but the goal of these projects is 1\% 
accuracy for $s_{13}$. For a baseline of L=10 km, with E$\simeq$ 4 MeV, we 
find a 35\% correction. Also, our SM method gives 
$P(\bar{\nu}_e\rightarrow \bar{\nu}_e)\neq 1.0$ even for
$s_{13}$=0.

 The CP violation probability (CPV), $\Delta P^{CP}_{\mu e}$, 
is strongly dependent on both of these important parameters.
After the determination of $s_{13}$, both the
JHF-Kamioka and Double Chooz projects might be able to determine the value 
of $\delta_{CP}$, since for most of the values of $\delta_{CP}$ these 
projects would have nearly a 1\% CPV.
No experiments are possible now, since beams of both neutrino and 
antineutrino with the same flavor would be needed, however, in the 
future such beams might be available.
Our results should help in planning such future experiments.
\clearpage

\begin{figure}[ht]
\begin{center}
\epsfig{file=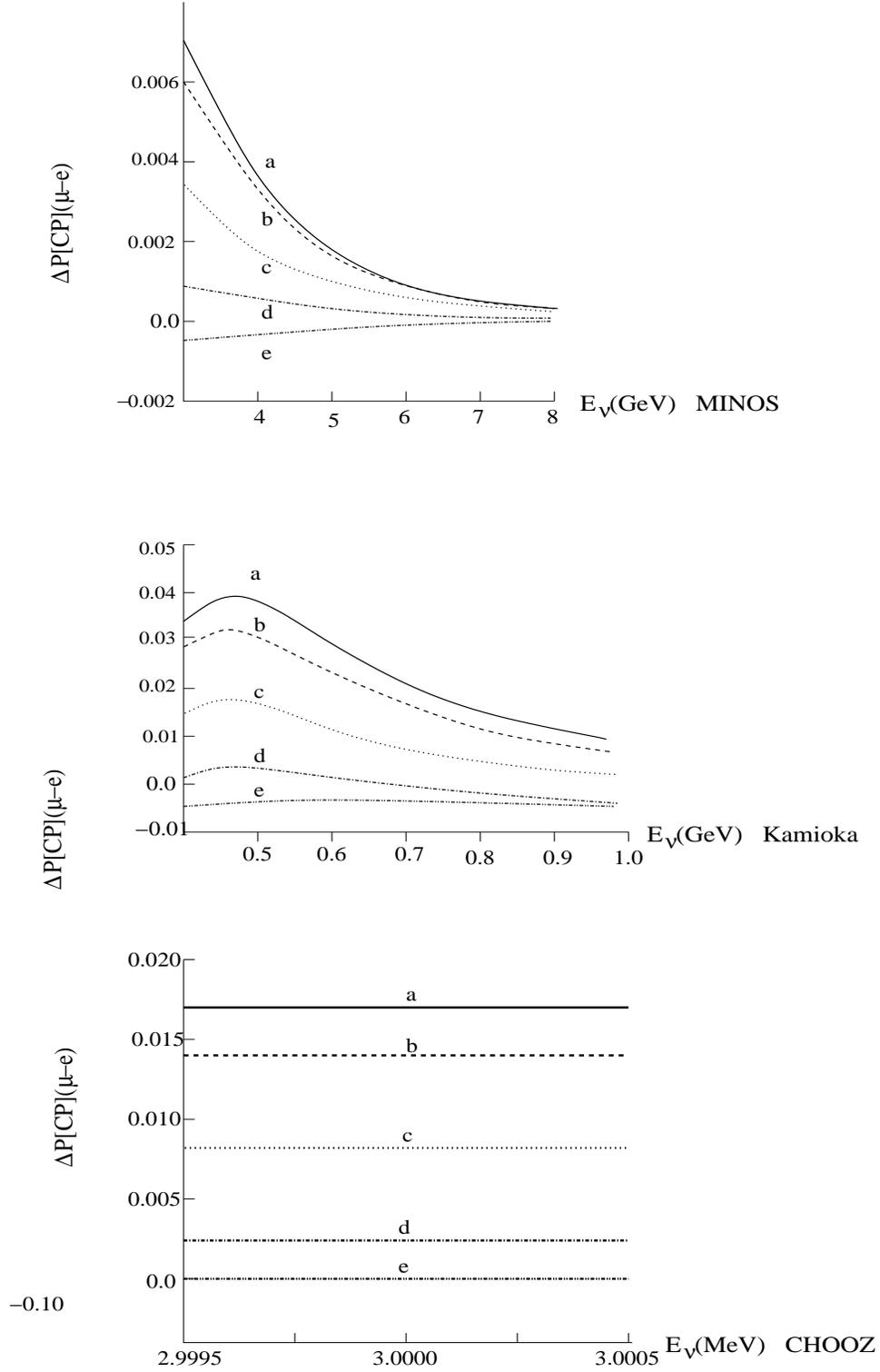,height=20cm,width=13cm}
\end{center}
\caption{The ordinate is $\Delta \mathcal{P}(\nu_\mu \rightarrow\nu_e)$ for 
MINOS(L=735 km), JHF-Kamioka(L=295 km), and CHOOZ(L=1 km).
s13=.19, and a, b, c, d, e for $\delta_{CP}$=$\pi/2$, $\pi/4$,
0.0, $-\pi/4$, $-\pi/2$}
\end{figure}

\clearpage

\begin{figure}[ht]
\begin{center}
\epsfig{file=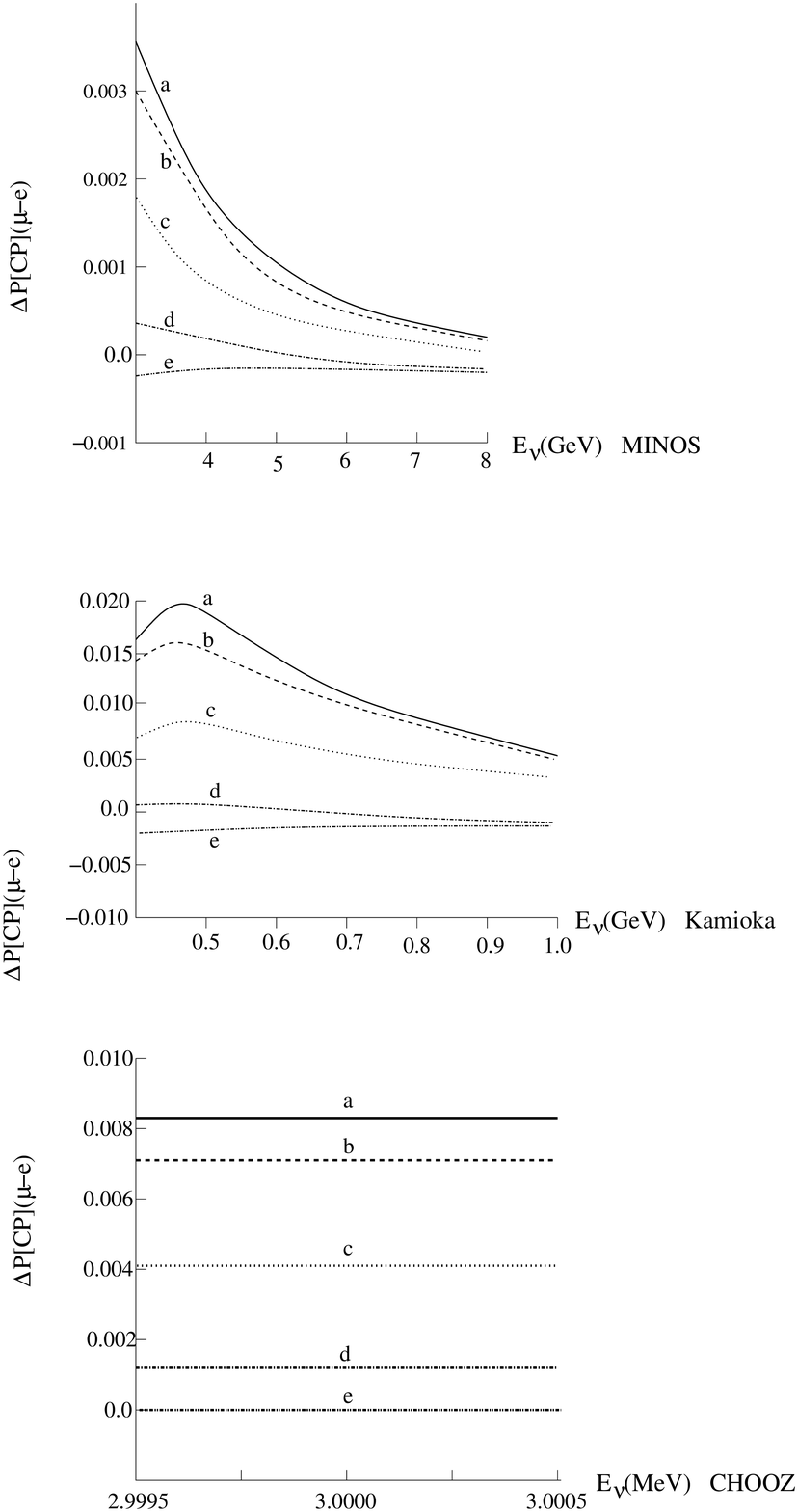,height=20cm,width=13cm}
\end{center}
\caption{The ordinate is $\Delta \mathcal{P}(\nu_\mu \rightarrow\nu_e)$ for 
MINOS(L=735 km), JHF-Kamioka(L=295 km), and CHOOZ(L=1 km).
s13=.095, and a, b, c, d, e for $\delta_{CP}$=$\pi/2$, $\pi/4$,
0.0, $-\pi/4$, $-\pi/2$}
\end{figure}

\clearpage
\section{Sterile Neutrinos and Pulsar Kicks}

  Pulsars are neutron stars with large magnetic fields spinning rapidly,
and therefore they emit light. It has been observed that they move with
large velocities, with the velocity increasing with luminoscity, L, as shown
in the figure below.
\vspace{8cm}

\begin{figure}[ht]
\begin{center}
\epsfig{file=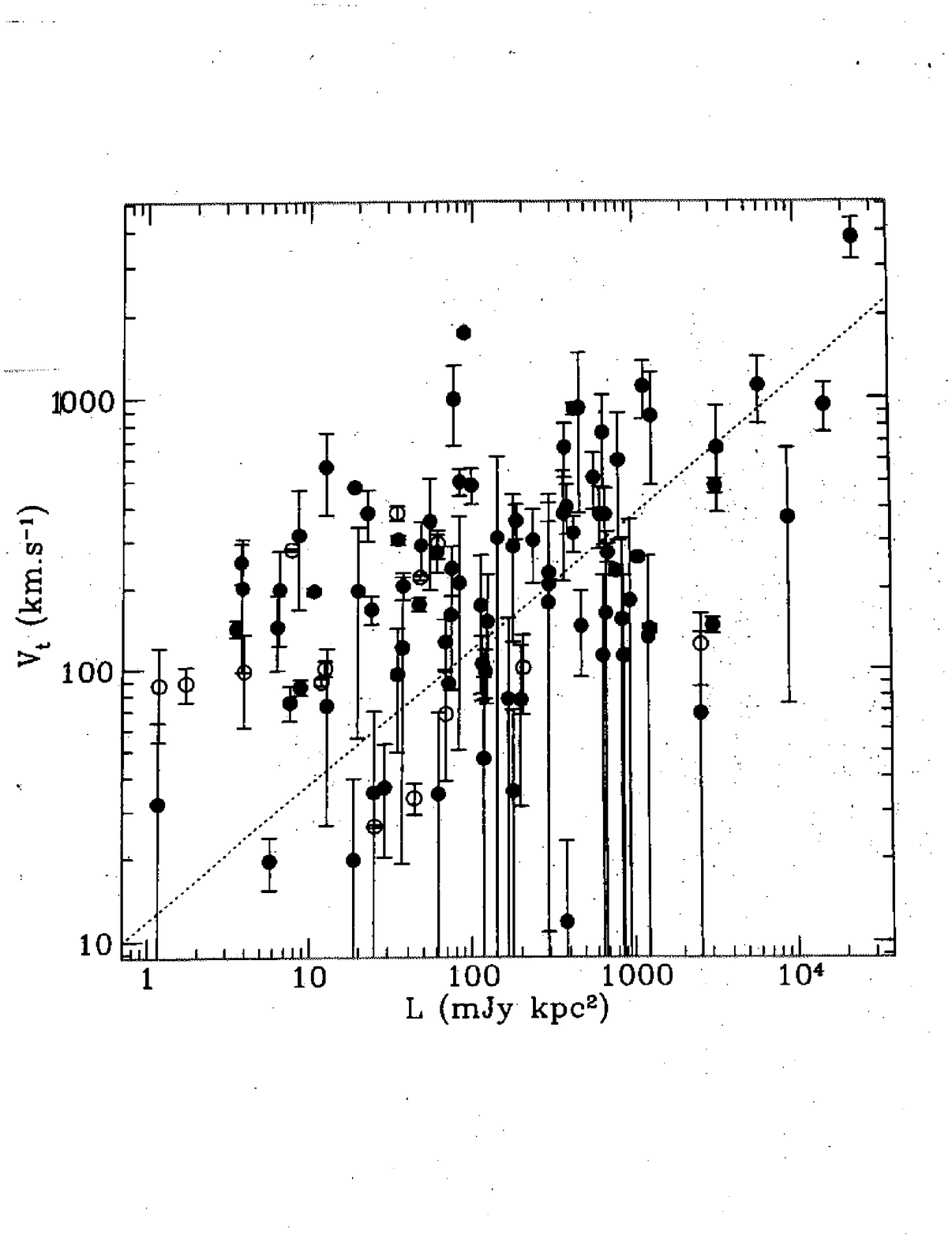,height=8cm,width=12cm}
\caption{Pulsar Speed vs. Luminoscity}
\end{center}
\end{figure}
\newpage
\subsection{Pulsar Kicks From Standard Neutrino Emission}

 The origin of these large velocities, called pulsar kicks, has been a problem
for decades. The pulsars are created during supernova explosions:
\vspace{1mm}

 A massive star ($\geq~$ 8 sun masses) burns its nuclear fuel in less than
a billion years and undergoes gravitational collapse:

\vspace{1mm}

  1. Collapse to density $> 10^{14}$ g cm$^{-3} >$ nuclear density

  2. Protoneutron star formed $\sim$ 0.01 s. 

\hspace{6mm}Neutrinos trapped in neutrinosphere. Radius of neutrinosphere 
$\sim$ 40 km.

 3. From 0.1 to 10 sec neutrinosphere contracts from $\sim$ 40 km to protostar
radius $\sim$ 10 km. The URCA process dominates neutrino emission (
$ n \rightarrow p +e^- \bar{\nu};  e^- + p \rightarrow n + \nu_e;
   e^+ + n \rightarrow p + \bar{\nu}_e$,
where the nucleons are polarized by the strong magnetic fields of the
protoneutron star. However, since the mean free path of neutrinos is
only 1 cm, and the asymmetric neutrinos are not emitted.
Therefore standard active neutrinos cannot explain pulsar kicks during the
first 10 sec.

4. From $\sim$ 10 to $\sim$ 50 sec n-n collisions dominate neutrino
production and protoneutron star cooling. 
 The modified URCA process dominates energy emmision by neutrinos.
\beq
           n + n &\rightarrow& n + p + e^- +\bar{\nu}_e \nonumber
\eeq

It was shown that the modified URCA process during 10-50s
might explain pulsar kicks\cite{hjk07}

\subsection{Pulsar Kicks From Sterile Neutrino Emission}

 Just as neutrinos of different flavors mix, as discussed above, active
neutrinos can oscillate into sterile neutrinos, neutrinos without an
even a weak interaction. 

 Sterile/active neutrino mixing  given by the mixing angle $\theta_m$

\beq
\label{sterileosc}
      |\nu_1> &=& cos\theta_m |\nu_e> -sin\theta_m |\nu_s> \\
      |\nu_2> &=& sin\theta_m |\nu_e> +cos\theta_m |\nu_s> \; . 
 \nonumber 
\eeq

At the time Ref\cite{hjk09} as published, $\theta_m$ was not known well
enough for the theory to be compared to measured pulsar velocities, but
recently\cite{abaz12} it has been determined to about 30 per cent: 
\beq
\label{sintheta} 
         sin^2(2\theta_m) &\simeq& 0.15 \pm .05 \; ,
\eeq 
and this was used to estimate the pulsar velocities\cite{kj12}.

 The neutrino emissivity, $e^\nu$=energy/(volume x $\Delta t$), where
$\Delta t$ is the time interval for the emission, from Ref\cite{fm} is
\beq
\label{6}
   e^\nu &=&\Pi^4_{i=1} \int \frac{d^3p^i}{(2\pi)^3}
\frac{d^3 q^\nu}{2 \omega^\nu(2 \pi)^3} 
 \int\frac{d^3q^e}{(2\pi)^3}
\nonumber \\
 && (2\pi)^4 \sum_{s_i,s^\nu} \frac{1} {2\omega^e_L}  \omega^\nu \mathcal{F}
|M|^2 \\
&&  \delta(E_{final}-E_{initial}) \delta(\vec p_{final}-
\vec p_{initial})  \nonumber \; ,
\eeq
where $M$ is the matrix element for the URCA process and
 $\mathcal{F}$ is the product of the initial and final Fermi-Dirac functions.
The $p^i$ are the two initial and two final nucleon momenta, and $q^\nu,q^e$
are the neutrino and electron momenta. 

The result for the asymetric neutrino emission along the direction of the
magnetic field, which can produce pulsar kicks, is\cite{hjk07}
\beq
\label{10}
  \epsilon^{AS} &\simeq& 0.64 \times 10^{21} T_9^7 P(0)\times f
\nonumber \\ 
&&{\rm erg\; cm^{-3}\; s^{-1}}=p_{ns} c \frac{1}{V_{eff} \Delta t} \; ,
\eeq
where $T_9 = T/(10^9 K)$, $p_{ns}$ is the neutron star momentum, $P(0)\simeq$
0.3 is 
the probability of the electron produced with the antineutrino
being in the lowest Landau state, f=.52 is the probability of the
neutrino being at the + z neutrinosphere surface~\cite{hjk07}, $V_{eff}$ is
the volume at the surface of the neutrinosphere from which neutrinos are
emitted, and $\Delta t \simeq 10 s$ is the time interval for the emission.

Although the sterile neutrino has no interaction, it oscillates back to
an electron neutrino as shown in Eq(\ref{sterileosc}). The effective mean
free path is abour five times longer than the active neutrinos. 
  $V_{eff}$, the volume at the surface of the neutrinosphere from which
neutrinos are emitted, is given by the mean free path of the sterile 
neutrino,
$\lambda_s$, and the radius of the neutrinosphere\cite{hjk09}:
\beq
\label{Veff}
  V_{eff} &=& (4\pi/3)( R_\nu^3-(R_\nu-\lambda_s)^3) \nonumber \\
          &\simeq& 4\pi R_\nu^2 \lambda/sin^2(2\theta_m) \; ,
\eeq
with $\lambda_s = \lambda/sin^2(2\theta_m)$, where $\lambda$ is the active
neutrino mean free path. Therefore
\beq
\label{pns}
    p_{ns}&=& 4\pi R_\nu^2 (\lambda/sin^2(2\theta)) (10 s)  \epsilon^{AS}
\; .
\eeq

 Using $p_{ns}=M_{ns} v_{ns}$, with $M_{ns}$the mass of the neutrino star, 
and taking $M_{ns}=M_{sun}=2 \times 10^{33}$ gm, one 
finds with $sin^2(2\theta)$=.15
\beq
\label{7}
      v_{ns} &\simeq& 22.3 \times 10^{-7} (\frac{T}{10^{10} K})^7 
\frac{km}{s} 
\; .
\eeq

During the early stages after the collapse of a massive star temperatures
T=20 MeV are expected\cite{fkmp03}. With T = 10 to 20 MeV the
pulsar velosities, with a 50\% range due to the uncertainty in 
$sin^2(2\theta)$,  are shown in the figure below\cite{kj12}.

\clearpage
\begin{figure}[ht]
\epsfig{file=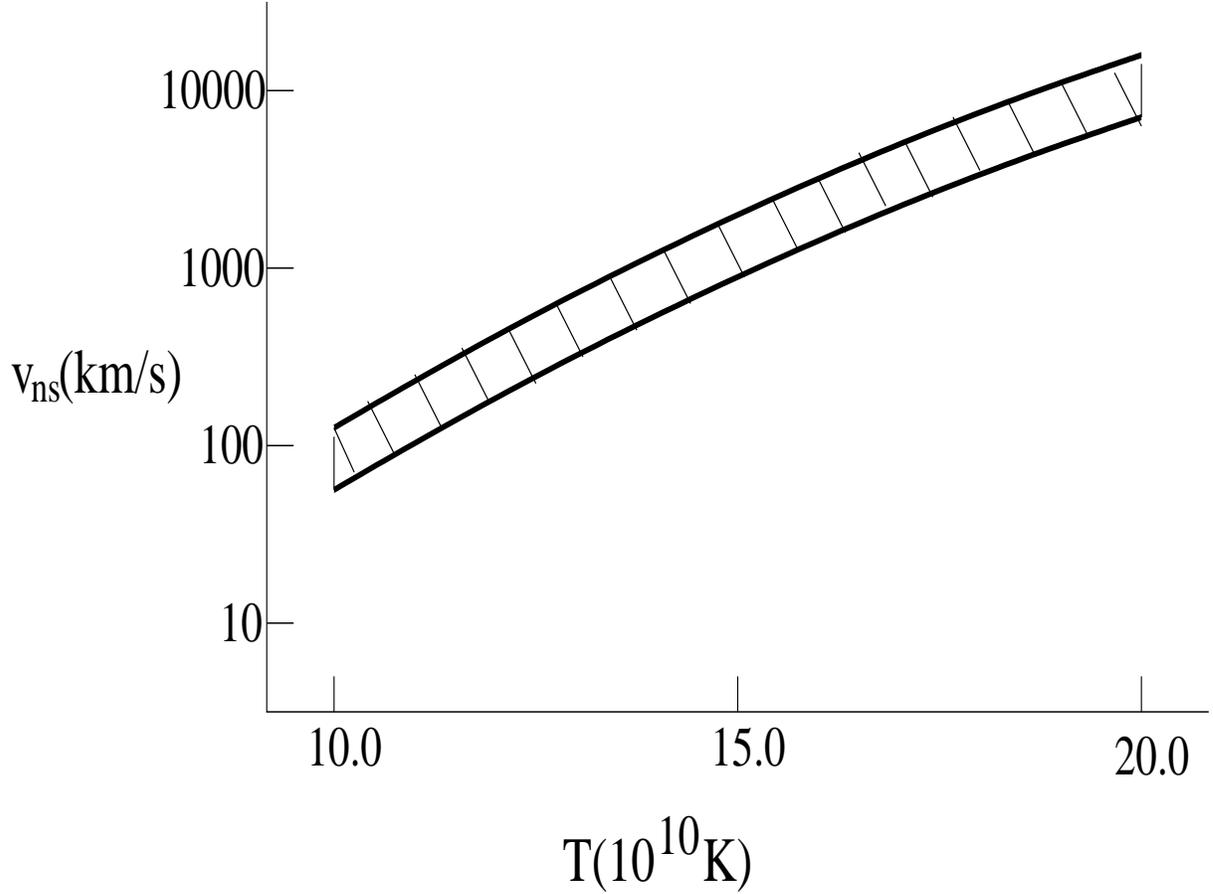,height=12cm,width=16cm}
\caption{The pulsar velocity as a function of T for $sin^2(2\theta)=.15 \pm 
.05$}
{\label{Fig.1}}
\end{figure}
\vspace{1.5cm}

As can be seen in Fig. 9., pulsar velocities of over 1000 km/s are 
predicted from sterile neutrino emission with the mixing angle recently 
measured\cite{abaz12}.  Therefore, sterile neutrino emission can account for
the large pulsar velocities for high luminosities (large T) 
that have been measured, as shown in Fig. 8.
This is a possible explanation of a puzzle that many have tried to explain 
for decades.
\newpage
\section{Neutrino Energies in a Neutrinosphere}

More than a decade ago, in preperation for studies of neutrino oscillations, 
energy eigenvalues of neutrinos in the earth were investigated by 
Freund\cite{freund01} using a cubic  eigenvalue formalism.

  Recently, this formalism has been used to find the energy eigenvalues of 
neutrinos in a neutrinosphere\cite{k13}, through which neutrinos must move to 
provide the pulsar kicks discussed in the previous section. We define the
energy of a neutrino with zero velocity as it's effective mass, which is
the definition of mass in vacuum.

 Using the method of Ref \cite{freund01}, the neutrino energy eigenvalues 
$E_i$,
\beq
                 H |\nu_i> &=& E_i |\nu_i> \nonumber \; ,
\eeq
are found as eigenvalues of the matrix $M$\cite{freund01} obtained from the
3 x 3 matrices $U$ and the Hamiltonian $H$:
\beq
 M=\left( \begin{array}{lcr}s_{13}^2 +\hat{A} + \alpha c_{13}^2s_{12}^2&\alpha
c_{13}s_{12}c_{12}&
s_{13}c_{13} -\alpha c_{13}s_{13}s_{12}^2\\ \alpha c_{13}s_{12}c_{12}& 
\alpha c_{12}^2&-\alpha s_{13}s_{12}c_{12}\\
s_{13}c_{13} - \alpha c_{13}s_{13}s_{12}^2&-\alpha s_{13}s_{12}c_{12}&c_{13}^2
 + \alpha s_{13}^2s_{12}^2 \end{array}
\right)
\eeq

With $c_{ij} \equiv cos(\theta_{ij}), s_{ij} \equiv sin(\theta_{ij})$, and
$\delta m_{rs}^2 \equiv m_r^2-m_s^2$, the parameters in M are:
$c_{12}^2 = 0.69, c_{13}^2 \simeq 1.0, \delta m_{12}^2 = 7.59 \times 10^{-5} eV^2,
\delta m_{13}^2 = 2.45 \times 10^{-3}, \alpha \equiv \delta m_{12}^2/\delta 
m_{13}^2 = 0.031, \hat{A} = 2E_\nu V/\delta m_{13}^2 $ ,
with $V$ the potential for neutrino interaction in matter. It is well-known 
that $V= \sqrt{2} G_F n_e$, where $G_F$ is the weak interaction Fermi constant,
and $n_e$ is the density of electrons in matter. For neutrinos in earth 
$V \simeq 1.13 \times 10^{-13} eV$, $\hat{A}\ll 1.0$

The eigenvalues of M satisfy the cubic equation (see Eq(17) in 
Ref\cite{freund01} with a=$-I_1, b=I_2, c=-I_3$):
\beq
\label{cubic}  
        \bar{E}_i^3+a \bar{E}_i^2 + b \bar{E}_i +c &=& 0 \\
         a &=& -(1+\hat{A} +\alpha) \nonumber \\
         b &=& \alpha + \hat{A} \alpha c_{12}^2 c_{13}^2 + \hat{A}(c_{13}^2
+\alpha s_{13}^2) \nonumber \\
         c &=& - \hat{A} \alpha c_{12}^2 c_{13}^2 \nonumber \; 
\eeq
with dimensionless quantities $\bar{E}_i = (E_i-E^0_1)/(E^0_3-E^0_1)$, where 
$E_i^0$ are neutrino energy eigenvalues with V=0.
We study neutrinos at rest, so $E_i \equiv m_i c^2 \equiv m_i$, with i = 1, 2, 
and 3; $E^0_i$ are the neutrino masses in vacuum, and $m_i$ are the effective
masses of neutrinos in matter. With the parameters given abovone finds,
\beq
\label{abc}
     a &=& -(1.031 + \hat{A}) \nonumber \\
     b &=& 0.031 + \hat{A} \nonumber \\
     c &=& -0.0209 \hat{A} \;.
\eeq

Taking  $E_\nu \simeq m_3 \simeq \sqrt{\delta m_{13}^2}$, as $m_1 \ll m_3$,
for which $\hat{A}$ is maximum, results in the largest matter effect on 
neutrino eigenstates. First, we solve the cubic equations for neutrinos
in vacuum.  From Eqs.(\ref{cubic},\ref{abc}) for V=0 ($\hat{A}=0$)
\beq
\label{Es}
       \bar{E}_1 &=& 3.08 \times 10^{-12} \simeq 0 \nonumber \\
       \bar{E}_2 &=& 0.031 \nonumber \\
       \bar{E}_3 &=& 1.0 \; ,
\eeq 
which is nearly the same as for neutrinos in earth ($\hat{A}\ll 1.0$).

The density of nucleons in the neutrinosphere is approximately that of atomic 
nuclear matter, $\rho_n =4 \times 10^{17} kgm/m^3$. Taking the ratio of the 
electron mass to the proton mass one finds for the electron density in the 
neutrinosphere $\rho_e \simeq 2 \times 10^{11}$ gm/cc, giving the neutrino 
potential in the neutrinosphere  $V \simeq 10^{-2}$ eV. 
.
This gives $\hat{A}_{ns}= 0.404$, which is $\hat{A}$ for neutrinos in a 
neutrinosphere.   Solving Eq(\ref{cubic}) with this dense matter potential 
one finds
\beq
\label{E2s}
       \bar{E}_1^{ns} &=& 0.0208 \nonumber \\
       \bar{E}_2^{ns} &=& 0.3998 \nonumber \\
       \bar{E}_3^{ns} &=& 1.0144 \; ,
\eeq

  Comparing Eq(\ref{E2s}) with Eq ({\ref{Es}), $m_3\simeq m_3(V=0)$,
while  $m_2 - m_1(V=0) \simeq 0.4 eV \simeq 13.0 \times (m_2(V=0)-m_1(V=0))$.
Therefore, the neutrino effective masses in the neutrinosphere are
quite different than in earth or vacuum. Although the three neutrino effective 
masses are approximately the same in earth matter as in vacuum, the large 
effect of matter on neutrino oscillations arise from a large baseline and 
depend on the energy of the neutrino beam. 

\section{Conclusions}
\vspace{5mm}

\hspace{1cm}{\bf 1) Although neutrino oscillations are promising for
determining TRV and CPV, new experimental facilities are needed.}
\vspace{5mm}

\hspace{6mm}{\bf 2) Neutrino oscillation, such as neutrino disappearance,
can provide accurate measurements of the parameters in U, the matrix
relating flavor neutrinos to mass neutrinos.}
\vspace{5mm}

\hspace{6mm}{\bf 3) Sterile neutrino emission during
the first 10 seconds after the gravitational collapse of a star can
explain the large pulsar velocities.}
\vspace{5mm}

\hspace{6mm}{\bf 4) The effective masses of neutrino in a neutrinosphere
are very different than those in earth of vacuum}
\vspace{1cm}

\Large{{\bf Acknowledgements}}\\
\normalsize
\vspace{2mm}

The author thanks Drs. M.B. Johnson and E.M. Henley for helpful discussions.
This work was supported in part by a grant from the Pittsburgh Foundation.

\end{document}